\documentstyle{article}
\renewcommand{\textheight}{236mm}

\renewcommand{\oddsidemargin}{-2cm}
\begin{document}

\begin{center}
\bf{ATOMIC DATA FOR PERMITTED RESONANCE LINES OF ATOMS AND IONS FROM H 
TO Si, AND S, Ar, Ca AND Fe}
\end{center}

\begin{center}
\vspace{1.5cm}
D. A. VERNER, E. M. VERNER, and G. J. FERLAND

\vspace{0.5cm}
Department of Physics and Astronomy, University of Kentucky,
Lexington, KY 40506, USA; verner@pa.uky.edy
\end{center}

\vspace{1.5cm}
We list vacuum wavelengths, energy levels, statistical weights, transition 
probabilities and oscillator strengths for permitted resonance spectral
lines of all ions of 18 astrophysically important elements (H through Si,
S, Ar, Ca, Fe). Using a compilation of experimental energy levels,
we derived accurate wavelengths for 5599 lines of 1828 ground-term multiplets
which have $gf$-values calculated in the Opacity Project. We
recalculated the Opacity Project multiplet $gf$-values to oscillator
strengths and transition probabilities of individual lines.
For completeness, we added 372 resonance lines of Ne~{\sc I}, Ar~{\sc I},
Fe~{\sc I} and Fe~{\sc II} ions which are not covered by the Opacity Project.
Intercombination and forbidden lines are not included in the present 
compilation.

\newpage
\begin{center}
\bf{INTRODUCTION}
\end{center}

\vspace{0.5cm}

Recently, a large international collaboration known as the Opacity
Project (hereafter OP; Seaton et al.$^1$) produced a complete set of accurate 
atomic data for 
all stages of ionization of astrophysically important elements:
$1 \leq Z \leq 14$ and $Z=16$, 18, 20, 26, where $Z$ is the 
atomic number. In particular, the OP data include 
oscillator strengths for all optically allowed transitions between 
states with $n \leq 10$ and $l \leq 4$, where $n$ and $l$ are the 
active-electron principal and orbital quantum numbers respectively. 
The OP data do not include intercombination and forbidden transitions.
The OP calculations are based on high level methods such as the R-matrix 
method (Burke, Hibbert, and Robb$^2$) and the new asymptotic techniques 
developed 
by Seaton$^3$. Extensive comparisons of the OP data with experiment 
and theory (Seaton et al.$^1$, Mendoza$^4$, Verner, Barthel, and Tytler$^5$, 
Verner 
and Yakovlev$^6$) have confirmed the high accuracy of the OP calculations. 
The OP data are available through the TOPbase database (Cunto et al.$^7$). 

However, the OP transition probabilities and oscillator strengths cannot
be directly used for spectroscopic applications because the OP gives all
the data for multiplets only, not for individual lines. Moreover,
the OP theoretical multiplet wavelengths are accurate to a few angstroms,
which is not good enough for spectroscopy. 

In an earlier paper (Verner et al.$^5$) we assigned accurate experimental
wavelengths to 1086 OP lines from the ground (zero energy) level at wavelengths
greater than 228~\AA. In the present paper, we extend this analysis for
all the OP permitted
resonance lines, i.e. lines connecting the ground level and excited 
fine-structure levels of the ground term to excited levels via optically
allowed transitions (transitions between levels of opposite
parity with $\left | \Delta L \right | \le 1$, 
$\left | \Delta S \right | =0$, $\left | \Delta J \right | \le 1$, where
$L$ is the orbital, $S$ the spin, 
and $J$ the total angular momentum quantum number).

We retrieved from the TOPbase 0.7 database the $gf$-values for all 
multiplets which involve the ground term.  These 6018 resonance multiplets 
include 23505 lines.  
In the OP calculations, relativistic effects are neglected, and $LS$-coupling
is assumed throughout. 
We calculated the $f$-values of all individual lines in 
multiplets based on $LS$-coupling rules (Russell$^8$).
To assign accurate wavelengths to these lines, we made a compilation of
experimental energy levels (complete to October 1995; references are given
in Table A). Vacuum 
wavelengths calculated
from experimental energy levels should be more reliable than measurements of
individual resonance lines, because the experimental energy levels are based
on the analysis of the whole experimental spectrum in each case. 
The number of digits we keep
in the tabulated wavelengths is determined by accuracy of the experimental 
energy levels quoted in the corresponding papers. For some hydrogenic ions,
Erikson$^{9}$ lists energy levels for various isotopes. In such cases, we 
used the values of energy levels of the most naturally abundant isotopes.

By this method, we assign accurate wavelengths to 5599 optically
allowed $LS$-type lines in 1828 multiplets (references to the OP
data are given in Table B). Beyond that, some gaps in the data still remain.
Firstly,
atomic data for Fe~{\sc I} and
Fe~{\sc II} ions are not included in the TOPbase
database. In the framework
of the OP extension known as the Iron Project (Hummer 
et al.$^{10}$),
Nahar$^{11}$ calculated transition probabilities for
21,589 allowed transitions of Fe~{\sc II} which have experimentally determined 
wavelengths. We included in our compilation all resonance lines of Fe~{\sc II}
from Nahar's paper. Multiplet tables for Fe~{\sc I},
which include 9759 lines, have been recently published
by Nave et al.$^{12}$. For our list, we selected the resonance lines
of  Fe~{\sc I} which have $gf$-values in the Nave et al. tables, and 
added the complete resonance multiplets of Fe~{\sc I} from the compilation by 
Morton$^{13}$. Secondly, the OP calculations
are based on $LS$-coupling, and only allowed $LS$-type lines are covered 
by our procedure. In particular, no data for lines of Ne~{\sc I} or Ar~{\sc I}
can be obtained from the Opacity Project. To fill this gap, we included the
resonance non-$LS$-type lines of Ne~{\sc I} and Ar~{\sc I} from the compilation
by Verner et al.$^5$.

In the whole, we list accurate wavelengths, energy levels, statistical weights,
transition probabilities and oscillator strengths for 5971 resonance lines.
Intercombination (i.e. $\left | \Delta S \right | > 1$) and forbidden lines 
are not included
in our compilation. 

We calculated also the multiplet energy levels by averaging the
weighted energy level values. The multiplet averaged wavelengths are
calculated by use of the averaged energy levels. For multiplets where
not all levels of the upper term are known, the averaged values are
obtained from the available levels. For these cases, the averaged multiplet
upper level
and wavelength are not the ``true'' averaged values.

We subsequently used accurate wavelengths to improve the $f$-values 
calculated by the OP.  The OP calculations
give line strength $S$ and transition energy $E$, where
$S$ is given by an integral over the atomic wave functions and does not 
involve $E$. Then $gf$-values are calculated from $S$ and $E$ since 
$gf = {\rm const} \times S \times E$ 
(or $gf = {\rm const} \times S/\lambda$). 
The OP used their theoretical values of $E$ to calculate $gf$-values. 
To improve the $gf$-values we now use the more accurate 
experimental wavelengths instead of the theoretical $E$ values. 
We recalculated the OP $f$-values as
\begin{equation}
f_{\rm improved} = f_{\rm OP} \times (\lambda_{\rm OP}/\lambda_{\rm exp}).
\end{equation}
Transition probabilities are calculated from oscillator strengths in
accordance with the standard formula:
\begin{equation}
A_{ki}=6.670 \times 10^{15}g_if_{ik}/(g_k \lambda ^2) \;\; {\rm s}^{-1},
\end{equation}
where $\lambda$ is in \AA.

To estimate the accuracy of the calculated oscillator strengths and
transition probabilities, we have compared the improved OP $f$-values
for iron ions with the $f$-values taken from the 
National Institute of Standards and Technology
(NIST) evaluated compilation
of atomic transition probabilities (Fuhr, Martin, and Wiese$^{14}$). 
Figure 1 shows
this comparison for all the NIST resonance iron lines with estimated
accuracy better than 10\%. All these lines are strong transitions of the
highly ionized (Fe~{\sc XV} and higher) ions.
The agreement is remarkably good. The rms
deviation of the improved OP $f$-values from the NIST $f$-values is 8.4\%,
and no values deviate by more than 15\%. Note that all the NIST values
are obtained by relativistic calculations, and relativistic effects
should be more important for iron than for other OP elements. 
Generally, for the strong lines which can be classified as $LS$-type 
allowed transitions, the improved OP data are fairly accurate. 
The obtained oscillator strengths and transition probabilities are
less accurate for weaker lines and for lines that are not pure $LS$-type,
in particularly, for high $l$ ($p-d$ and higher) transitions.
We do not list here branching ratios which indicate the purity of
$LS$ coupling, since such data are available for a smaller portion of
the lines only. However, special caution should be taken when using 
the listed oscillator strengths and transition probabilities for
the lines of neutrals and first ions of the third row elements
(Mg, Al, Si, S, Ar), especially for the weak lines.

The main incompleteness
of our compilation comes from the fact that all lines which
cannot be classified as allowed $LS$-type lines are not covered here
(except Ne~{\sc I} and Ar~{\sc I} lines).
However, the present compilation is accurate and complete for the permitted
$LS$-type lines of all ionization states of 18 elements. This compilation
is supplementary to the compilation of the resonance absorption lines above
912~\AA by Morton$^{13}$, the compilation of ground level absorption lines 
above 228~\AA by
Verner et al.$^5$, and the NIST evaluated compilations of the atomic
transition probabilities by Martin et al.$^{15}$, Fuhr et al.$^{14}$,
Wiese et al.$^{16}$).

Tables I and II in electronic form 
are available through anonymous ftp at 
asta.pa.uky.edu, directory
dima/lines, or through the World Wide Web page ``Atomic Data for 
Astrophysics'',\\
http://www.pa.uky.edu/$\sim$verner/atom.html.

\vspace{0.7cm}
{\it Acknowledgements}.  This work was supported by grants from the National
Science Foundation (AST 93-19034) and NASA (NAGW 3315).
We are grateful to M. J. Seaton for useful discussions on the Opacity Project 
data. We are also grateful to the Opacity Project team for their efforts
to make the data available before publication through the TOPbase database.
We acknowledge the use of the TOPbase 0.7 database
installed by A. K. Pradhan at the Ohio State University. We thank K. T. Korista
for helpful conversations. 
We are indebted to both referees for suggestions which improved the
presentation of the tables.

\vspace{1cm}
{\it References}

\vspace{0.5cm}
\noindent
1. M. J. Seaton, C. J. Zeippen, J. A. Tully, et al.,
Rev. Mex. Astron. Astrofis. {\bf 23,} 19 (1992)\\
2. P. G. Burke, A. Hibbert, and W. D. Robb, J. Phys. B {\bf 4,} 153 (1971)\\
3. M. J. Seaton, 1985, J. Phys. B {\bf 18,} 2111 (1985)\\
4. C. Mendoza, in: P. L. Smith and W. L. Wiese (eds.), 
{\it Atomic and Molecular Data for Space Astronomy: Needs, Analysis and 
Availability,} Lecture Notes in Physics, {\bf 407,}
85 (Springer-Verlag, Berlin, 1992)\\
5. D. A. Verner, P. D. Barthel, and D.  Tytler, Astron. Astrophys. Suppl. 
{\bf 108,} 287 (1994)\\
6. D. A. Verner and D. G. Yakovlev, Astron. Astrophys. Suppl. {\bf 109,} 125 
(1995)\\
7. W. Cunto, C. Mendoza, F. Ochsenbein, and C. J. Zeippen, Astron. Astrophys.
{\bf 275,} L5 (1993)\\
8. H. M. Russell, Astrophys. J. {\bf 83,} 129 (1936)\\
9. G. W. Erikson, J. Phys. Chem. Ref. Data {\bf 6,} 831 (1977)\\
10. D. G. Hummer, K. A. Berrington, W. Eissner, et al., Astron. Astrophys.
{\bf 279,} 298 (1993)\\
11. S. N. Nahar, Astron. Astrophys. {\bf 293,} 967 (1995)\\
12. G. Nave, S. Johansson, R. C. M. Learner, A. P. Thorne, and J. W. Brault,
Astrophys. J. Suppl. {\bf 94,} 221 (1994)\\
13. D. C. Morton, Astrophys. J. Suppl. {\bf 77,} 119 (1991)\\
14. J. R. Fuhr, G. A. Martin, and W. L. Wiese, J. Phys. Chem. Ref. Data 
{\bf 17,} Suppl. 4 (1988)\\
15. G. A. Martin, J. R. Fuhr, and W. L. Wiese, J. Phys. Chem. Ref. Data 
{\bf 17,} Suppl. 3 (1988)\\
16. W. L. Wiese, J. R. Fuhr, and T. M. Deters, J. Phys. Chem. Ref. Data,
Monograph 7, in press (1995)\\
17. W. C. Martin, J. Phys. Chem. Ref. Data {\bf 2,} 257 (1973)\\
18. I. Johansson, Ark. Fys. {\bf 15,} 169 (1959)\\
19. R. Crossley, J. Opt. Soc. Am. B {\bf 1,} 266 (1984)\\
20. L. Johansson, Ark. Fys. {\bf 23,} 119 (1963)\\
21. L. Johansson, Ark. Fys. {\bf 20,} 489 (1961)\\
22. B. L\"{o}fstrand, Physica Scripta {\bf 8,} 57 (1973)\\
23. G. A. Odintzova and A. R. Striganov, J. Phys. Chem. Ref. Data {\bf 8,} 63 
(1979)\\
24. A. \"{O}lme, Physica Scripta {\bf 1,} 256 (1970)\\
25. A. \"{O}lme, Ark. Fys. {\bf 40,} 35 (1969)\\
26. M. Eidelsberg, J. Phys. B {\bf 7,} 1476 (1974)\\
27. C. E. Moore, NSRDS-NBS {\bf 3,} Section 3 (1970)\\
28. L. Engstr\"{o}m, P. Bengtsson, C. Jup\'{e}n, and M. Wesrelind, J. Phys. B 
{\bf 25,} 2459 (1992)\\
29. C. E. Moore, NSRDS-NBS {\bf 3,} Section 5 (1975)\\
30. K. B. S. Eriksson, Physica Scripta {\bf 28,} 593 (1983)\\
31. C. E. Moore, NSRDS-NBS {\bf 3,} Section 4 (1971)\\
32. A. M. Malvezzi, Physica Scripta {\bf 27,} 413 (1983)\\
33. C. E. Moore, NSRDS-NBS {\bf 3,} Section 7 (1976)\\
34. W. C. Martin, V. Kaufman, and A. Musgrove, J. Phys. Chem. Ref. Data 
{\bf 22,} 1179 (1993)\\
35. S. G. Pettersson, Physica Scripta {\bf 26,} 296 (1982)\\
36. C. E. Moore, NSRDS-NBS {\bf 3,} Section 10 (1983)\\
37. C. E. Moore, NSRDS-NBS {\bf 3,} Section 9 (1980)\\
38. C. E. Moore, NSRDS-NBS {\bf 3,} Section 8 (1979)\\
39. R. C. Isler, C. Jup\'{e}n, and I. Martinson, Physica Scripta {\bf 47,} 32 
(1993)\\
40. K. Liden, Ark. Fys. {\bf 1,} 229 (1949)\\
41. H. P. Palenius, Ark. Fys. {\bf 39,} 15 (1969)\\
42. H. P. Palenius, Physica Scripta {\bf 1,} 113 (1970)\\
43. C. E. Moore, NSRDS-NBS {\bf 35,} Atomic Energy Levels Vol. {\sc I} (1971)\\
44. L. Engstr\"{o}m, Physica Scripta {\bf 31,} 379 (1985)\\
45. L. Engstr\"{o}m, Physica Scripta {\bf 29,} 113 (1985)\\
46. U. Feldman, G. A. Doschek, D. J. Nagel, W. E. Behring, and R. D. Cowan, 
Astrophys. J. {\bf 187,} 417 (1974)\\
47. W. Persson, Physica Scripta {\bf 3,} 133 (1971)\\
48. W. Persson, Phys. Rev. A {\bf 43,} 4791 (1991)\\
49. S. Goldsmith and A. S. Kaufman, Proc. Phys. Soc. {\bf 81,} 544 (1963)\\
50. H. Hermansdorfer, J. Opt. Soc. Am. {\bf 62,} 1149 (1972)\\
51. K. Bockasten, R. Hallin, and T. P. Hughes, Proc. Phys. Soc. {\bf 81}, 522 
(1963)\\
52. G. Tondello and T. M. Paget, J. Phys. B. {\bf 3}, 1757 (1970)\\
53. N. J. Peacock, R. J. Speer, and M. G. Hobby, J. Phys. B. {\bf 2}, 798 
(1969)\\
54. W. C. Martin and R. Zabulas, J. Phys. Chem. Ref. Data {\bf 10,} 153 
(1981)\\
55. W. C. Martin and R. Zabulas, J. Phys. Chem. Ref. Data {\bf 9,} 1 (1980),\\
56. V. Kaufman and W. C. Martin, J. Phys. Chem. Ref. Data {\bf 20,} 83 (1991)\\
57 W. C. Martin and R. Zabulas, J. Phys. Chem. Ref. Data {\bf 8,} 817 (1979),\\
58. V. Kaufman and W. C. Martin, J. Phys. Chem. Ref. Data {\bf 20,} 775 
(1991)\\
59. S. Khardi, M. C. Buchet-Poulizac, J. P. Buchet, M. Carr\'{e}, A. Denis, J.
D\'{e}sesquelles, A. E. Livingston, S. Martin, and Y. Ouerdane, Physica 
Scripta {\bf 49,} 571 (1994)\\
60. W. C. Martin and R. Zabulas, J. Phys. Chem. Ref. Data {\bf 12,} 323 
(1983)\\
61. A. E. Khramida and E. Tr\"{a}bert, Physica Scripta {\bf 51,} 209 (1995)\\
62. W. C. Martin, R. Zabulas, and A. Musgrove, J. Phys. Chem. Ref. Data 
{\bf 19,} 821 (1990),\\
63. V. Kaufman and W. C. Martin, J. Phys. Chem. Ref. Data {\bf 22,} 279 
(1993)\\
64. P. Bengtsson, C. Jup\'{e}n, L. Engstr\"{o}m, A. Redfors, and
M. Westerlind, Physica Scripta {\bf 48,} 413 (1993)\\
65. W. Whaling, W. H. C. Anderson, M. T. Carle, J. W. Brault, and H. A. Zarem, 
J. Quant. Spectr. Radiat. Transfer {\bf 53,} 1 (1995)\\
66. L. Minnhagen, J. Opt. Soc. Am. {\bf 61,} 1257 (1971)\\
67. J. E. Hansen and W. Persson, J. Phys. B {\bf 20,} 693 (1987)\\
68. F. Bredice, M. Gallardo, J. G. Reyna Almandos, A. G. Trigueros,
and C. J. B. Pagan, Physica Scripta {\bf 51,} 446 (1995)\\
69. L. W. Phillips and W. L. Parker, Phys. Rev. {\bf 60,} 301 (1941)\\
70. A. J. J. Raassen, L. C. Snoek, H. Volten, P. H. M. Uylings, F. G. Meijer, 
and Y. N. Joshi, Astron. Astrophys. Suppl. {\bf 95,} 223 (1992)\\
71. M. Raineri, F. Bredice, M. Gallardo, J. G. Reyna Almandos, C. J. B. Pagan,
and A. G. Trigueros, Physica Scripta {\bf 45,} 584 (1992)\\
72. B. C. Fawcett, Physica Scripta {\bf 30,} 326 (1984)\\
73. P. Bengtsson, L. Engstr\"{o}m, and C. Jup\'{e}n, Physica Scripta {\bf 49,}
297 (1994)\\
74. B. C. Fawcett, A. H. Gabriel, and T. W. Paget, J. Phys. B {\bf 4,} 986 
(1971)\\
75. W. A. Deutschman and L. L. House, Astrophys. J. {\bf 144,} 435 (1966)\\
76. W. A. Deutschman and L. L. House, Astrophys. J. {\bf 149,} 451 (1967)\\
77. B. C. Fawcett and R. W. Hayes, Physica Scripta {\bf 8,} 244 (1973)\\
78. J. P. Connerade, N. J. Peacock and R. J. Speer, Solar Phys. {\bf 18,} 63 
(1971)\\
79. G. E. Bromage and B. C. Fawcett, Mon. Not. Roy. Astr. Soc. {\bf 178,} 605 
(1977)\\
80. K. G. Widing, Astrophys. J. {\bf 197,} L33 (1975)\\
81. G. D. Sandlin, G. E. Brueckner, V. E. Scherrer, and R. Tousey, Astrophys. 
J. {\bf 205,} L47 (1976)\\
82. J. Sugar and C. Corliss,  J. Phys. Chem. Ref. Data {\bf 14,} Suppl. 2 
(1985)\\
83. J. O. Ekberg, R. Smitt, B. Skogvall, and A. Borgstr\"{o}m, Physica Scripta
{\bf 41,} 217 (1990)\\
84. J. O. Ekberg, Astron. Astrophys. Suppl.{\bf 101,} 1 (1993)\\
85. T. Shirai, Y. Funatake, K. Mori, J. Sugar, W. L. Wiese, and T. Nakai, J. 
Phys. Chem. Ref. Data {\bf 19,} 127 (1990)\\ 
86. J. A. Fernley, K. T. Taylor, and M. J. Seaton, J. Phys. B 
{\bf 20,} 6457 (1987)\\
87. G. Peach, H. E. Saraph, and M. J. Seaton, J. Phys. B {\bf 21,} 
3669 (1988)\\
88. J. A. Tully, M. J. Seaton, and  K. A. Berrington, J. Phys. B
{\bf 23,} 3811 (1990)\\
89. D. Luo and A. K. Pradhan, J. Phys. B {\bf  22,} 3377 (1989)\\
90. A. Hibbert and M. P. Scott, J. Phys. B {\bf 27,} 1315 (1994)\\
91. K. Butler, C. Mendoza, and C. J. Zeippen, J. Phys. B {\bf 26,} 4409
(1993)\\
92. C. Mendoza, W. Eissner, M. Le Dourneuf, and C. J. Zeippen, 
J. Phys. B {\bf 28,} 3485 (1995)\\
93. S. N. Nahar and A. K. Pradhan, J. Phys. B {\bf 26,} 1109 (1993)\\
94. P. M. J. Sawey and K. A. Berrington, J. Phys. B {\bf 25,} 1451 (1992)\\
95. H. E. Saraph, P. J. Storey, and K. T. Taylor, J. Phys. B 
{\bf 25,} 4409 (1992)\\
96. B. C. Fawcett, B. B. Jones, and R. Wilson, Proc. Phys. Soc. {\bf 78,} 
1223 (1961)\\
97. V. Kaufman and L. Minnhagen, J. Opt. Soc. Am. {\bf 62,} 92 (1972)\\
98. E. S. Chang, W. G. Schoenfeld, E. Biemont, P. Quinet, P. Palmeri, 
Physica Scripta {\bf 49,} 26 (1994)\\
99. L. Minnhagen, J. Opt. Soc. Am. {\bf 63,} 1185 (1973)\\
100. S. Johansson, private communication (1994)

\newpage
\begin{center}
\bf{EXPLANATION OF TABLES}
\end{center}

\vspace{0.5cm}
{\bf Table I. Wavelengths and Oscillator Strengths of Permitted Resonance 
Absorption Lines}\\
The data are sorted by element from H to Fe, ionization state 
(starting with neutral state), multiplet (starting with the largest
averaged multiplet wavelength), and wavelength (decreasing wavelength
within the multiplet). Theoretical energy levels are given in
square brackets. The averaged multiplet data values of the listed levels
and wavelengths are given in
{\it italic}. 

\vspace{0.3cm}
\begin{tabular}{ll}
Species & Atom or ion \\
Transition & Electronic configurations of lower (ground term) and upper 
             levels.\\
 &If an upper level has an uncertain identification in the Opacity Project 
  (OP),\\
 & the relevant configuration is marked by semicolon (:)\\
 & If the order of levels within a spectroscopic series in the OP does not \\
 & coincide with the order of levels within the spectroscopic series stated 
 in the\\ 
 &corresponding source of experimental energy levels, the OP identifications\\
 & are used, and the relevant configurations are marked by asterisk (*)\\
Multiplet & Terms of lower and upper levels\\
$\lambda$ & Vacuum wavelength in \AA\\
$E_i$, $E_k$ & Energies of lower and upper levels in cm$^{-1}$\\
 & If an upper level energy has a question mark in the source of\\
 & experimental energy levels, we also indicate it and the corresponding\\
 & wavelength by question mark (?)\\
 & If experimental spectroscopy of the upper term is not complete, and\\
 & the experimental energy of an upper level is unknown, we leave the\\
 & upper level energy and the corresponding wavelength entries blank\\
$g_i$, $g_k$ & Statistical weights of lower and upper levels\\
$A_{ki}$ & Atomic transition probability in s$^{-1}$\\
$f_{ik}$ & Absorption oscillator strength 
\end{tabular}

\vspace{0.5cm}
{\bf Table II. Finding List of Lines}\\
The lines are sorted by increasing wavelength.

\vspace{0.3cm}
\begin{tabular}{ll}
$\lambda$ & Vacuum wavelength in \AA\\
Species & Atom or ion \\
$g_i$, $g_k$ & Statistical weights of lower and upper levels\\
$f_{ik}$ & Absorption oscillator strength 
\end{tabular}

\vspace{1cm}
\begin{center}
\bf{FIGURE CAPTION}
\end{center}

Figure 1. Improved Opacity Project oscillator strengths (Eq. 1) vs. the 
oscillator
strengths from Fuhr et al.$^{14}$ for iron lines which are
classified by Fuhr et al.$^{14}$ as accurate to within 10\%.

\end{document}